\documentstyle[prl,aps,multicol,epsf]{revtex}
\renewcommand{\narrowtext}{\begin{multicols}{2} \global\columnwidth20.5pc}
\renewcommand{\widetext}{\end{multicols} \global\columnwidth42.5pc} 

\begin{document}

\newcommand{\be}{\begin{equation}}
\newcommand{\ee}{\end{equation}}
\newcommand{\bea}{\begin{eqnarray}}
\newcommand{\eea}{\end{eqnarray}}
\newcommand{\nt}{\narrowtext}
\newcommand{\wt}{\widetext}

\title{Ghost excitonic insulator transition in layered graphite}
\author{D. V. Khveshchenko}
\address{Department of Physics and Astronomy, University of North
Carolina, Chapel Hill, NC 27599}
\maketitle

\begin{abstract}
Some unusual properties
of layered graphite, including a linear energy dependence of the quasiparticle damping and 
weak ferromagnetism at low doping, are explained as a result of  
the proximity of a single graphene sheet to the excitonic insulator phase
which can be further stabilized in a doped system of many layers stacked in the
staggered ($ABAB\dots$) configuration.
\end{abstract}

\nt
The continuing interest in carbon-based materials 
has intensified both experimental and theoretical 
efforts to understand their electronic properties and interaction-driven transitions 
in these systems. 

The problem of electronic instabilities in a single sheet of graphite     
has been previously studied by means of the Hartree-Fock and density functional   
methods with a focus on the short-ranged (including on-site and nearest neighbor) 
Hubbard-like repulsive interactions, and the inter-site repulsion was found to favor 
a charge density wave (CDW) ground state \cite{hf}.
This analysis did not provide, however, a proper account of the Coulomb forces
that remain long-ranged due to the lack of conventional screening in semimetals, 
such as graphite.

In contrast, the authors of Ref.\cite{guinea} specifically 
focused on the role of the Coulomb interactions.
From the renormalization group calculation in the second order 
in the dimensionless Coulomb coupling $g=2\pi e^2/\varepsilon_0v$  they
concluded that the renormalized coupling monotonously decreases 
at low energies and, therefore, it can not 
cause any instability of the gapless paramagnetic ground state of graphite.

Besides, the authors of Ref.\cite{guinea} suggested a possible explanation for the 
experimentally observed \cite{xu} linear energy dependence 
of the quasiparticle damping (defined as the imaginary part of the quasiparticle dispersion
$\epsilon=E_{\bf p}+i\Sigma(\epsilon)$ ) which they found to behave 
as $\Sigma(\epsilon)\sim\epsilon/\ln^2\epsilon$ at low energies.
However, the large estimated value of the bare coupling constant 
$(g\gtrsim 10)$ calls these results into question and warrants further investigation.

In the present Letter, we revisit the problem of the Coulomb 
interacting electrons in layered graphite and study 
the nature of the ground state and the quasiparticle spectrum at strong coupling. 
This time around, we employ a non-perturbative approach by solving a non-linear 
equation for the electron Green function, which will allow us to 
ascertain the status of the previous results obtained in perturbation theory 
and test our theoretical predictions against several pieces of the existing
experimental evidence.

The semimetallic energy band structure of a single graphene sheet 
gives rise to the conduction and valence bands' touching
each other in the two inequivalent  
$K$-points located at the corners of the hexagonal two-dimensional (2D) Brillouin zone.
In the absence of interactions, the low-energy quasiparticle excitations with the momenta
in the vicinity of these points labeled as $i=1,2$ 
have linear dispersion $E^{(0)}_{\bf p}=\pm v{\bf p}$,
the velocity $v$ being proportional to the 
width of the electronic $\pi$-band $t\approx 2.4 eV$ \cite{semenoff}. 

These excitations can be formally described by a pair of two-component (Weyl) spinors 
$\psi_{i\sigma}$, each carrying a spin index $\sigma$, 
which are composed of the Bloch states residing on the two 
different sublattices of the bi-partite hexagonal lattice of the graphene sheet. 
In what follows, we choose to combine them into one four-component Dirac spinor $\Psi_\sigma=
(\psi_{1\sigma}, \psi_{2\sigma})$ and also treat the number of the spin components $N$ 
as an adjustable parameter, the physical case corresponding to $N=2$.

The use of the Dirac spinor representation
allows one to cast the free quasiparticle Hamiltonian in the relativistic-like form
where $v$ is playing the role of the speed of light 
\be
H_0=iv\sum_{\sigma =1}^N \int_{\bf r}{\overline \Psi}_\sigma
({\hat \gamma}_1\nabla_x+{\hat \gamma}_2\nabla_y)\Psi_\sigma
\label{kin}
\ee
where ${\overline \Psi}_\sigma=
\Psi^\dagger_\sigma{\hat \gamma}_0$ and the reducible representation 
of the $4\times 4$ $\gamma$-matrices 
${\hat \gamma}_{0,1,2}=(\tau_{3}, i\tau_{2}, -i\tau_{1})\otimes\tau_3$ 
given in terms of the triplet of the Pauli matrices 
$\tau_{i}$ satisfies the usual anticommutation relations: 
$\{{\hat \gamma}_{\mu},{\hat \gamma}_{\nu}\}=2 {\rm diag}(1, -1, -1){\bf 1}\otimes{\bf 1}$.

In the four-spinor representation,
the electron Coulomb interaction reads as  
\be
H_C={v\over 4\pi}\sum^{N}_{\sigma,\sigma^\prime=1}\int_{{\bf r},{\bf r}^\prime}
{\overline \Psi}_{\sigma}({\bf r}){\hat \gamma}_0\Psi_{\sigma}
({\bf r}){g\over {|{\bf r}-{\bf r}^\prime|}}
{\overline \Psi}_{\sigma^\prime}({\bf r}^\prime)
{\hat \gamma}_0\Psi_{\sigma^\prime}({\bf r}^\prime)
\label{int}
\ee 
Despite the apparent lack of the Lorentz invariance, 
Eq.(2) remains invariant under arbitrary $U(2N)$ rotations
of the $2N$-component vector $(\Psi_{L\sigma}, \Psi_{R\sigma})$
composed of the chiral Dirac fermions defined as:  
$\Psi_{L,R\sigma}={1\over 2}({\bf 1}\pm{\hat \gamma}_5)\Psi_\sigma$, 
where the matrix ${\hat \gamma}_5=
{\bf 1}\otimes{\tau_2}$ anticommutes with any ${\hat \gamma}_\mu$. 

The chiral invariance of Eqs.(1,2)  
brings about the possibility of spontaneous chiral symmetry breaking (CSB),
akin the phenomenon that has long been studied in the 
relativistic fermion theories. The CSB transition
manifests itself in the appearance of a fermion mass and gapping of the fermion spectrum,
thus breaking the continuous chiral symmetry from $U(2N)$ down to $U(N)\otimes U(N)$
and developing a non-zero expectation value   
$<\sum_{\sigma}^N{\overline \Psi}_\sigma({\bf r})\Psi_\sigma({\bf r})>=
 <\sum_{\sigma}^N(\psi^\dagger_{\sigma}(A)\psi_{\sigma}(A) 
- \psi^\dagger_{\sigma}(B)\psi_{\sigma}(B))>$. The latter 
corresponds to the electron density modulation  
which alternates between the two sublattices ($A$ and $B$).

In light of the above, one can identify the CSB order parameter with the site-centered CDW
and thus relate it to the $p=0$ value of the gap 
function $\Delta_p$ appearing in the renormalized (and, generally, non-Lorentz
invariant) fermion Green function 
\be
{\hat G}_p=Z_p[(\epsilon{\hat \gamma}_0 - v_p{\vec p}{\hat {\vec \gamma}})+\Delta_p]^{-1}
\label{G0}
\ee
where the interaction effects can also give rise to the non-trivial    
wave function ($Z_p$) and velocity ($v_p/v$) renormalization factors.

Because of its intrinsically non-perturbative nature, the phenomenon of CSB   
evades weak-coupling analysis based on perturbation theory.
Nonetheless, similar to its relativistic counterpart \cite{app1},
the CSB can be revealed by a non-perturbative solution of the system of non-linear 
equations for the fermion Green function 
(hereafter ${\hat p}=\epsilon{\hat \gamma}_0 - v{\vec p}{\hat {\vec \gamma}}$)
\be
{\hat G}^{-1}_p={\hat p}+\int\!{d^3{k}\over (2\pi)^3}
\Gamma_{p,k}{\hat \gamma}_0{\hat G}_{p+k}{\hat \gamma}_0V_k, 
\label{G}
\ee
vertex function $\Gamma_{p,k}$, and effective Coulomb interaction 
$V_k=1/[({\bf q}/gv)+ N\chi(\omega,{\bf q})]$
which gets strongly modified by the intra-layer polarization of the Dirac fermions 
\be
\chi_k={\rm Tr}
\int {d^3{p}\over (2\pi)^3}\Gamma_{p,k}
{\hat \gamma}_0{\hat G}_{p+k}{\hat \gamma}_0{\hat G}_{p}
\label{chi}
\ee
A further analytical progress is hindered
by the fact that, as a result of the interaction's $V_k$ being explicitly 
non-Lorentz invariant,
the gap function $\Delta_p$ can feature separate dependencies on the energy $\epsilon$ 
and momentum $\bf p$ variables. 
Therefore, we choose to proceed directly with the finite temperature 
counterpart of Eq.(4), in which case the Lorentz invariance
is broken regardless of the symmetry of the fermion interactions.

In order to get a preliminary insight into the problem 
we resort to the same approximations as those made in the previous studies of 
CSB in the context of $QED_3$ \cite{app1}.
To this end, we first neglect the wave function, velocity,
and vertex renormalizations ($Z_p=v_p/v=\Gamma_{p,k}=1$)
in Eq.(4) whose scalar part then becomes a closed equation for the fermion gap function.  
As shown in \cite{app1}, neglecting the above renormalizations in the gap equation
suffices for 
establishing the existence of its non-trivial solution(s) 
and estimating a critical value $N_c$ of the only remaining free  
parameter, the number of fermion species.

Taking the sum over the discrete Matsubara frequencies 
we then arrive at the momentum-dependent gap equation 
\bea
\Delta_{\bf p}=
\int {d^2{\bf k}\over 8\pi^2}
\frac{\tanh E_{\bf k}/2T}{E_{\bf k}}
\frac{\Delta_{\bf k}}{|{\bf k}-{\bf p}|/gv + N\chi(0,{\bf k}-{\bf p})}
\label{gapeq}
\eea 
where $E_{\bf p}={\sqrt {v^2{\bf p}^{2}+\Delta_{\bf p}^2}}$. 
Next, we approximate the exact finite temperature fermion 
polarization (\ref{chi}) by that computed in the massless case. 
By doing so, we overestimate the contribution 
of the fermion momenta ${\bf p}\lesssim \Delta_{\bf p}$ which 
is, however, unimportant, as long as the gap remains much smaller than the 
high-momentum cutoff $\Lambda$ comparable to the maximum span of the Brillouin zone.
As shown below, this condition is indeed satisfied for $N$ close to the critical value $N_c$. 

At $\Delta_{\bf p}=0$ the fermion polarization is given by the approximate formula
\bea 
\chi(0,{\bf q})
={2T\over \pi v^2_F}\int^\infty_0 dx
\ln [2\cosh({v{\bf q}\over 2T}{\sqrt {x(1-x)}})]\nonumber\\
\approx {1\over 8v^2}[v{\bf q}+cT\exp(-{v{\bf q}\over cT})] 
\label{Pi}
\eea
which, for $c=16\ln 2/\pi$, provides an up to a few percent 
accurate interpolation between the two opposite limits:
$v{\bf q}\gg T$ where Eq.(\ref{Pi}) agrees with the zero-temperature 
result $\chi_k\propto {\sqrt {k^2}}$ and $v{\bf q}\ll T$ where it exhibits thermal 
screening $\chi_0\propto T$ \cite{aitchison}. 

Notably, at strong coupling ($g\gg 1$) the screened Coulomb interaction $V_k$ 
becomes independent of the bare coupling constant   
and assumes a universal form $V_k\approx 1/\chi(0,{\bf k})$  
governed by the fermion polarization (7).

Upon differentiating Eq.(6) with respect to the momentum $\bf p$
one finds that for $E_{\bf p}>T$ this non-linear integral equation
reduces to a linear differential one 
\be
{d^2\Delta_{\bf p}\over d{\bf p}^2}+{2\over {\bf p}}{d\Delta_{\bf p}
\over d{\bf p}}+{2\over \pi N}{\Delta_{\bf p}\over {\bf p}^2}=0
\label{diff}
\ee
which has to be supplemented by the boundary conditions $\Delta_{0} < \infty$ and 
$(\Delta_{\bf p}+{\bf p}d\Delta_{\bf p}/d{\bf p})|_{{\bf p}=\Lambda}=0$.

In turn, Eq.(\ref{diff}) can be readily identified with the radial Schroedinger equation  
for the $s$-wave zero energy level in the potential
that behaves as $\propto 1/{\bf p}^2$ for ${\bf p}>T/v$. From 
the textbook solution of this problem \cite{LL} we infer that for $N>N_c=8/\pi$ 
there are two independent solutions
$\Delta^{\pm}_{\bf p}\propto 1/{\bf p}^{(1\pm\sqrt{1-N_c/N})/2}$, 
neither of which can satisfy 
the above boundary conditions. In contrast, for $N<N_c$ there exists a 
solution with infinitely many nodes, consistent with the infinite number
of the negative energy levels in the $1/{\bf p}^2$-potential. 
Thus, only in this "centerward downfall" regime does the solution 
\be
\Delta_{\bf p}\sim {T^{3/2}\over \sqrt{v{\bf p}}}
\sin({1\over 2}\sqrt{{N_c\over N}-1}\ln{v{\bf p}\over T})
\ee
monotoneously decrease in the interval 
$T/v<{\bf p}<\Lambda$ and satisfy the boundary condition
at ${\bf p}=\Lambda$ which reads as 
\be
\sqrt{{N_c\over N}-1}\ln{v\Lambda\over T}=2\pi n -
2\tan^{-1}({1\over 2}\sqrt{{N_c\over N}-1})
\label{NT}
\ee
where $n$ is a positive integer.
We note, in passing, that due to the formal similarity between the underlying equations,
a solution with the similar critical properties 
has also been discovered in the context of 
the 2D Cooper pairing near the antiferromagnetic instability \cite{Chubukov}.
 
The highest critical temperature $T_c(N)$ below which 
the CSB order parameter sets in corresponds to $n=1$:   
\be
T_c(N)\approx v\Lambda\exp(-{2\pi\over \sqrt{{N_c/N}-1}}) 
\label{T_c}
\ee
According to Eq.(11), at non-zero temperatures the critical number of fermion species 
gets reduced: $N_c(0)-N_c(T)\approx 4\pi^2 N_c(0)/\ln^2(v\Lambda/T)$.

In the case of $QED_3$, the progressively more and more refined 
numerical simulations \cite{aitchison} have 
demonstrated that the results of the original analytical 
approach of Ref.\cite{app1} which yielded a solution similar to Eq.(9)   
remain robust against relaxing the above approximations
and taking into account both the wave function renormalization and
the vertex function satisfying the Ward identity $\Gamma_{p,0}=Z_p$. 

Likewise, a numerical analysis of the coupled Eqs.(4) and (5) 
confirms the existence of the solution (9) in a whole domain bordered 
by the critical line (11) in the $N-T$ plane \cite{leal}.

Also, the numerically evaluated 
characteristic ratio $2\Delta_0/T_c\approx 10$ appears to be close to that
found in Ref.\cite{aitchison} which is  
substantially greater than the BCS value corresponding to the
solution $\Delta_{\bf p}=const$ of the gap equation with a 
momentum-independent kernel. 

As far as the nature of the CSB transition is concerned, the observed $N$-dependence of the
zero-momentum fermion gap $\Delta_0\propto\exp(-2\pi/\sqrt{N_c/N-1})$ 
prompts one to identify the breaking of the continuous chiral symmetry as a   
topological (Kosterlitz-Thouless-type) phase transition 
in $2+1$ dimensions \cite{app1}. Thus, the finite temperature CSB transition  
occurs between the two phases which are both chirally symmetrical,
and therefore
a bosonic Goldstone mode must be present in the quasi-ordered phase. 

As regards the critical number of fermion species itself, 
the recent symmetry-based argument made in the context of $QED_3$  
shows that the gap equation systematically overestimates the actual value 
of $N_c$ which may, in fact, be as low as $3/2$ \cite{app2} while 
the gap equation yields $N_c^{QED}=32/\pi^2$ \cite{app1}. 

The demonstrated formal relationship between the finite temperature
$QED_3$ and the problem of a single graphene sheet 
suggests that in the latter case
the actual critical number of fermion species might also be 
less than two, hence no CSB occurs.

Nonetheless, even at $N>N_c$
the nearby CSB transition, albeit unreachable at any $g$, 
can still have a profound effect on the 
quasiparticle spectrum both above and below the crossover
into the quantum-critical regime associated 
with the zero-temperature quantum-critical 
point at $N_c$. 

In the quantum disordered (low-temperature) 
regime $T \lesssim T^\star(N)$, where the crossover temperature $T^\star(N)$
vanishes at $N\to N_c{+0}$ in the same manner as $T_c(N)$ given by Eq.(11)
for $N\to N_c{-0}$, the only solution of Eqs.(4) and (5)
is a massless fermion propagator which exhibits a 
suppression of the residue of the bare quasiparticle pole:     
$Z_p\to 0$ for ${p}\to 0$, while the velocity $v_p$ undergoes singular renormalization 
and monotonously increases with decreasing momentum,
unlike in the Lorent-invariant $QED_3$ where it remains constant. 

By contrast, in the quantum-critical regime $T \gtrsim T^\star(N)$
the fermion propagator features a simple
pole, whereas the temperature- (but no longer momentum-)
dependent factors $Z_T$ and $v_T$ control 
its residue and the effective velocity, respectively.
The fermion damping is then determined by the self-consistent equation  
\bea
\Sigma(\epsilon)={1\over N}
\int{d\omega d{\bf q}\over (2\pi)^3} 
\left\{\tanh\frac{\varepsilon+\omega}{2 T}-
\coth\frac{\omega}{2 T}\right\}\times
\nonumber\\
{\rm Im}\left[{\varepsilon+\omega+i\Sigma(\varepsilon+\omega) \over
(\varepsilon+\omega+i\Sigma(\varepsilon+\omega))^2-v^2{\bf q}^2}\right]
\!{\rm Im}{1\over \chi(\omega,{\bf q})}
\eea
\label{gamma}
which yields the universal solution: $\Sigma(\epsilon)\sim{\rm max}(\epsilon, T)$,
in a general agreement with the time-resolved two-photon photoemission data 
taken in the energy range $0.4 < \epsilon < 2 eV$ \cite{xu}.
One can expect that this characteristic 
signature of the CSB-related quantum-critical behavior will
be even more pronounced in the case of a graphite monolayer 
deposited on an insulating surface, whereas a conducting substrate   
would hamper the possibility of observing the linear damping
due to strong metallic screening.

The predicted linear damping should be possible to observe 
in angular-resolved photoemission which can 
specifically probe the vicinity of the $K$-points. 
On the contrary, the angular-averaged data are going to be affected
by such details of the graphite bandstructure as, e.g., the saddle point 
in the quasiparticle dispersion which occurs at $\epsilon\approx 1.5 eV$
if the momentum resides at one of the $M$-points of the Brillouin zone.
This saddle point was recently argued to be a likely cause of the additional plateau-like feature  
observed in the angular-averaged $\Sigma(\epsilon)$ \cite{Moos} 
which is, therefore, unrelated to the many-body phenomena discussed in this Letter.

In a stack of graphite layers with the inter-layer spacing 
$d$, the screened intra-layer Coulomb interaction remains dominated by the 
polarization $\chi_k$ only at ${\bf q}>1/Ngd$. At still lower momenta 
the kernel in Eq.(\ref{gapeq}) becomes less singular ($V_k\propto 1/{\sqrt {\bf q}}$), 
thus reducing the range of temperatures and/or energies where the
quantum-critical behavior associated with the nearby CSB transition 
can be observed in electron photoemission.

A finite inter-layer hopping $t_\perp\approx 0.27 eV$
provides another cutoff below which the particle-hole pairing 
correlations cease to drive the system towards the opening of the excitonic gap, 
and the quasiparticle damping becomes quadratic in energy for $\epsilon\lesssim t_\perp$.

In contrast, the inter-layer Coulomb interaction has the opposite effect of 
nudging a stack of graphite
layeres closer to the CSB instability. To elucidate this point, 
we recall that the common form of graphite has a crystal structure 
of well-separated hexagonal layers stacked in a staggered ($ABAB\dots$) configuration.
As a result, each layer gets naturally divided into two sublattices 
formed by the atoms positioned just above and below the centers and corners  
of the hexagons in the two adjacent layers, respectively.

Thus the inter-layer Coulomb repulsion strengthens the system's propensity towards
developing the CDW instability by favoring spontaneous depletion of one of the
two sublattices (accompanied by excess occupation of the complementary one)
which alternates between the layers in order 
to keep the electrons in the neighboring layers as far apart as possible. 

Formally, one can incorporate this effect of the inter-layer Coulomb 
interaction into the effective single-layer description by adding a short-ranged 
four-fermion term $\lambda(\sum_{\sigma}^N{\overline \Psi}_\sigma
\Psi_\sigma)^2$ into Eq.(2). With such a term present,   
the CSB transition occurs for any number of fermion species, including $N>N_c$, 
provided that the strength of this 
coupling exceeds a certain critical value ($\lambda > \lambda_c(N)$)
which grows with $N$ \cite{leal}. 

By further elaborating on the solution of Eq.(6) obtained in the physical case $N=2$
we also find out that, upon doping the system of graphite layers, 
the excitonic insulating ground state tends to spontaneously
develop a non-zero spin polarization  $<\psi^\dagger_\sigma(A)\psi_\sigma(A)
+\psi^\dagger_\sigma(B)\psi_\sigma(B)> \sim \mu\delta_{\sigma\sigma_0}$
proportional to the chemical potential $\mu$ introduced by doping \cite{leal}.
 
This observation sheds light on the possible origin of  
weak ferromagnetism which was recently
observed in highly oriented pyrolitic graphite (HOPG) \cite{Kopelevich}.
Our fundings suggest that the latter 
might be not that different from the mechanism proposed in 
the recent studies of hexaborides believed to be 3D excitonic insulators \cite{rice}.  
Notably, the authors of Ref.\cite{Kopelevich} excluded magnetic impurities as a
possible cause of the ferromagnetic behavior of the magnetization hysteresis loops
(also consistent with the electron spin resonance data) 
observed in the samples showing the insulator-like temperature dependence of the resistivity.
Elaborating on the analysis of Ref.\cite{rice} we predict that  
if the excitonic instability proved to be at work, 
the weak ferromagnetism would have to disappear above a certain level of doping
corresponding to the chemical potential $\mu_c\sim\Delta_0$. 

In summary, we study the problem of the Coulomb interaction-driven electronic instabilities 
in layered graphite and propose a new explanation for the experimentally observed
linear quasiparticle damping which might be a result of the relative proximity 
of a single graphene sheet to 
the zero-temperature quantum-critical point corresponding to the transition to the
2D excitonic insulator. In lightly doped layered 
graphite, the excitonic instability give rise to the formation
of the site-centered CDW ground state 
exhibiting weak ferromagnetism, as observed experimentally.
Together with the recently proposed explanation \cite{DVK} of the 
apparent semimetal-insulator transition in applied magnetic field \cite{Kempa}
as a phenomenon of the magnetic field-driven CSB, it lends a further support to the  
discovered formal relationship between the problem of layered graphite
and the behaviors found in the relativistic theories of the 2D Dirac fermions.

The author is grateful to Y. Kopelevich for communicating his experimental results prior
to publication and S. Washburn for a discussion. 
This research was supported by the NSF under Grant No. DMR-0071362.

\wt
\end{document}